\documentclass{elsart}
\usepackage{graphicx}
\begin{document}
\begin{frontmatter}
\title{Shuttle Instability in Self-Assembled Coulomb Blockade 
Nanostructures}
\author[address1]{A. Isacsson}, 
\author[address1,address2]{L. Y. Gorelik},
\author[address1,address3]{M. V. Voinova},
\author[address1]{B. Kasemo} 
\author[address1]{R. I. Shekhter} \and
\author[address1]{M. Jonson}
\address[address1]{Department of Applied Physics, Chalmers University 
of Technology and  G{\"o}teborg University, S-412 96 G{\"o}teborg, Sweden}
\address[address2]{B. Verkin Institute for Low Temperature Physics and 
Engineering, 310164 Kharkov, Ukraine}
\address[address3]{Kharkov State University, 310077 Kharkov, Ukraine}
\begin{abstract}
We study a simple model of a self-assembled, room temperature Coulomb-blockade 
nanostructure containing a metallic nanocrystal or grain connected by soft 
molecular links to two metallic electrodes. Self-excitation of periodic grain 
vibrations at 10 - 100 GHz is shown to be possible 
for a sufficiently large bias voltage leading to a novel `shuttle mechanism' 
of discrete charge transfer and a current through the nanostructure 
proportional
to the vibration frequency. For the case of weak electromechanical coupling an
analytical approach is developed which together with Monte Carlo
simulations shows that the shuttle instability for structures with 
high junction 
resistances leads to hysteresis in the current - voltage characteristics. 
\end{abstract}
\end{frontmatter}

\section{Introduction}
Conventional microelectronics is approaching a limit where further 
miniaturization is no longer possible. This has motivated a vigorous search
for alternative technologies such as `single electronics', which is based
on Coulomb charging effects in ultrasmall structures \cite{one,L}. 
A novel approach to building such structures --- nature's own approach --- 
is self-assembly using molecular recognition processes to form complex 
functional units. Familiar examples of molecules with this ability are
amphiphilic chain molecules (e.g. thiol) in solution which form ordered 
films on easily polarizable metal surfaces, antibodies which find and bind
to specific molecular targets, and DNA strands which recognize and bind 
to matching sequences. Recent progress include the successful use of 
self-assembled DNA templates for making tiny silver wires connecting 
macroscopic gold electrodes \cite{Sivan} and the demonstration of 
room-temperature 
Coulomb blockade behavior in novel composite mesoscopic structures 
containing both metallic elements and self-assembled organic matter 
\cite{two,three,four}. Charging effects in the latter structures are in 
the focus of the work we present here.

\begin{figure}[htbp]
  \begin{center}
   \leavevmode
\includegraphics[width=0.6\linewidth]{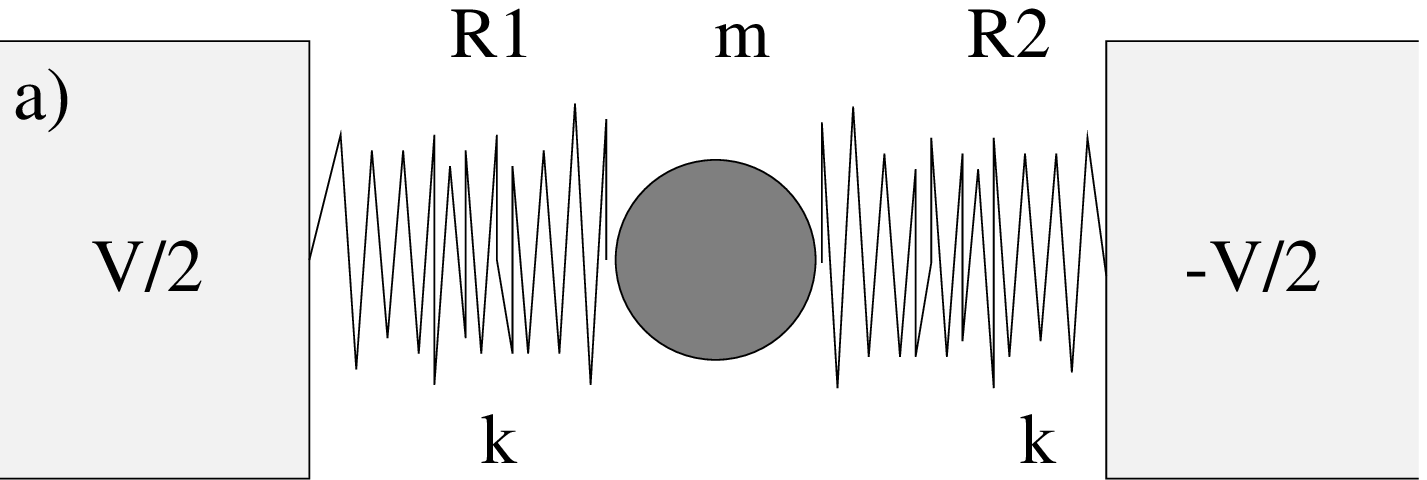}
\includegraphics[width=0.6\linewidth]{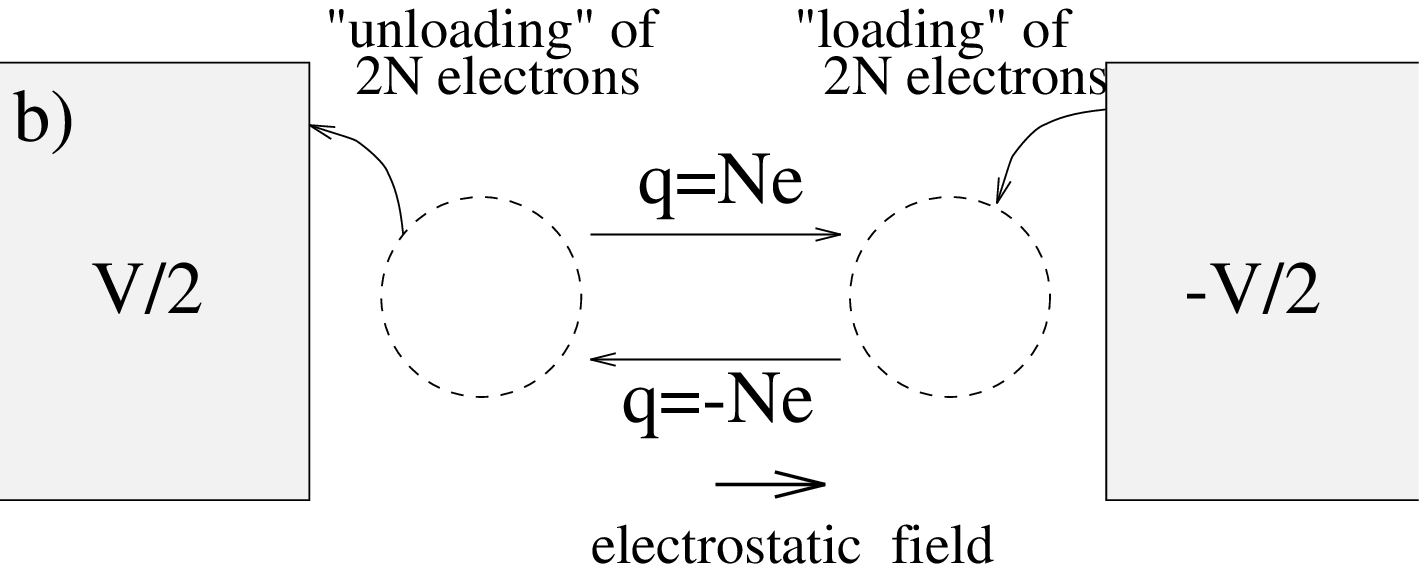}
\bigskip
    \caption{(a) Simple model of a soft Coulomb blockade system in which a 
  metallic grain (center) is linked to two electrodes by elastically
  deformable 
  organic molecular links. (b) Dynamic instabilities occur since in the
  presence of a sufficiently large
  bias voltage $V$ the grain is accelerated by the same
  electrostatic force towards first one, then the other electrode. A cyclic
  change in direction is caused by the repeated ``loading'' of  electrons
  near the negatively biased electrode and the subsequent ``unloading'' of the
      same at the positively biased electrode. 
      As a result the sign of the net grain charge
      alternates leading to an oscillatory grain motion and a novel ``electron
      shuttle" mechanism for charge transport.
      }
    \label{fig:one}
 \bigskip
  \end{center}
\end{figure}

The crucial aspect of the new room-temperature Coulomb blockade structures 
from the point of view of our work is that they contain metallic grains or 
molecular clusters with a typical size of 1-5 nm that can vibrate; their 
positions are not necessarily fixed. This is because the dielectric material 
surrounding them is elastic and consists of mechanically soft organic 
molecules. These molecular inter-links have elastic moduli which
are typically two or three orders of magnitude smaller than those of 
ordinary solids \cite{Mm}. Their
ohmic resistance is high and of order $10^7$ - $10^8$ ohm, while
at the same time they are extremely small --- a few nanometers in size.
A large Coulomb blockade effect in
combination with the softness of the dielectric medium implies that 
charge transfer may give rise to a significant deformation of these 
structures as they
respond to the electric field associated with a bias voltage.
Hence the position of a grain with respect to, for instance,
bulk metallic leads is not necessarily
fixed. For the model system of Fig.~\ref{fig:one} --- containing one 
metallic cluster connected by molecular links to two metallic electrodes
--- we have recently shown \cite{prl}
that self-excitation of mechanical grain vibrations at 10 - 100 GHz
accompanied by barrier deformations is possible for a 
sufficiently large bias voltage. This effect amounts to a
novel `shuttle mechanism' for electron transport. 

The purpose of the present paper is to carry the analysis presented in
Ref.~\cite{prl} quite a bit further. Strictly speaking the analytical
part of the analysis in
our earlier work is valid for low tunnel-barrier resistances when
the rate of charge redistribution between grain and leads, 
inversely proportional to the tunneling resistance, is assumed
to be so large in comparison with the vibration frequency that the 
stochastic fluctuations in grain charge during a single vibration period
are unimportant. In order to
describe the opposite limit of low charge redistribution frequencies 
characteristic of high-resistance tunnel barriers we develop 
here a new approach for 
considering the coupling betwen charge fluctuations and grain vibrations. 
We show that in the case of weak electromechanical coupling
(i.e. when a change of the grain charge by $e$ does not significantly
affect the  vibrational motion) the main
results which can be obtained within the 
model introduced in Ref.~\cite{prl} can be proven to be correct even for 
high resistance junctions. In this high ohmic limit a new scenario for the
shuttle instability becomes possible leading to hysteresis in the
current - voltage characteristics. 

The paper is organized as follows: In Section 2 we present our model
composite Coulomb blockade system as well as a derivation of
the equations governing the charge transport through the system.
Then, in Section 3 we derive the dynamical equations
for the case when the `fast' grain oscillations have been averaged
out so that only the `slow' variation in oscillation amplitude remains.
In Section 4 these equations are used to analyze the shuttle 
instability and the resulting current - voltage characteristics in our 
model Coulomb blockade system. The results are in good agreement with 
our earlier numerical results. The analytically solvable model
is in addition useful for analyzing the nature of the loss of stability 
as the shuttle mechanism sets in. Such an analysis --- supplemented by 
Monte Carlo simulations --- is carried out in 
Section 5, where we find that depending on the resistances of the tunnel 
junctions the transition from the static regime to the shuttle regime
can be associated either with smooth increase in the amplitude of the
self-oscillations or with a jump in the amplitude at the transition.
Adopting nomenclature from the theory of oscillations \cite{Andronov}
we are dealing with either soft or hard excitation of 
self-oscillations,
where hard excitation is associated with a hysteretic behavior of the
current as the bias voltage is swept up and 
down\footnote{In the language of phase transitions --- taking
the oscillation amplitude to be the order parameter --- the `soft'
case corresponds to a second order transition and the `hard' case to a first
order transition}. 
Finally, in Section 6 we present our conclusions.

\section{Model System}
In this Section we present a model of
the simplest possible composite Coulomb blockade system that retains the
properties of interest for us. It 
consists of one small metallic grain of mass $M$
connected by elastic molecular links to two bulk leads on either side,
as shown in Fig.~1a.
The electrostatic potential of the grain $\phi$ is a linear function of the 
grain charge $Q$ and the bias voltage ${V}$
\begin{displaymath}
\phi=\frac{Q}{C(X)}+a(X){V} \, ,
\end{displaymath}
where $X$ denotes the displacement of the grain from the equilibrium 
position in the centre of the system. This relation defines the capacitance
$C(X)$ that appears in the theory.
In a symmetric situation, as the one considered in this paper, 
$\phi(Q,{V},X)=\phi(Q,-{V},-X)$ and hence the coefficients $C(X)$ 
and $a(X)$ are even and odd 
functions of $X$ respectively.
For small displacements we have $\phi\approx Q/C$, where $C=C(0)$.
The magnitude of $C$ corresponds typically to the size of the grain.
There are three different forces acting on the grain; a linear elastic 
restoring
force $F_{el}=-kX$, a dissipative damping force 
$F_d=-\gamma_{d}\dot{X}$
and an electrostatic force $F_{q}=\frac{d}{dx}\mathcal{F}$,
\begin{displaymath}
  \mathcal{F}\equiv U_\mathrm{e}-\frac{{V}} 
{2}(Q_\mathrm{L}-Q_\mathrm{R}) \, ,
\end{displaymath}  
where the electrostatic energy $U_\mathrm{e}$ of the system and the charges 
in the left (right) lead $Q_\mathrm{L}$ ($Q_\mathrm{R}$) are 
considered to be functions of $Q$ and ${V}$. The function $\mathcal{F}$ is 
bilinear in $Q$ and ${V}$ hence
\begin{displaymath}
  \mathcal{F}=\frac{Q^2}{2C(X)}+\vartheta(X)Q{V}+
\frac{{V}^2}{2}b(X) \, .
\end{displaymath}
For a symmetric junction 
\begin{displaymath}
  \mathcal{F}(Q,{V},X)=\mathcal{F}(Q,-{V},-X)
\end{displaymath} 
implying that $\vartheta(X)$ is an odd function and that $C(X)$ 
and $b(X)$ are even.
Then, for small displacements,  
\begin{displaymath}
 F_{q}=\frac{{V}}{L}Q,\mbox{ } L^{-1}\equiv\frac{d\vartheta(X)}{dX}{\Bigg|}_{X=0}
\end{displaymath}
and one can consider $\mathcal{E}\equiv{V}/L$ as an effective electrostatic 
field 
induced by the bias voltage acting on the grain between the leads. This way 
we arrive 
at the estimate $L\sim${\em distance between the leads}.

Knowing the forces that act on the grain, we may
now introduce its equation of motion,
\begin{equation}
 M\ddot{X}+\gamma_{d}\dot{X}+kX=\mathcal{E}Q(t) \, .
 \label{e.EOM}
\end{equation}
If the Coulomb charging energy, $U_c=e^2/C$, satisfies 
$U_c\gg \hbar/RC,\beta^{-1}$ 
where $R$ is the characteristic tunneling resistance and $\beta$ is the 
inverse temperature,
one can expect a strong quantization of the charge $Q(t)$ in units of the 
elementary charge $e$,
\begin{displaymath}
Q(t)=en(t) \, ,
\end{displaymath}
where $n(t)$ will be a step function which can take on only integer 
values.
Changes in $n(t)$ with time are due to quantum transitions of electrons 
between the grain and the leads.  According to the `orthodox' Coulomb 
blockade theory \cite{L}
the probability for the transition 
\begin{displaymath}
  (n,Q_\mathrm{L,R})\rightarrow(n\pm1,Q_\mathrm{L,R}\mp e)
\end{displaymath}
to occur during a small time interval $\Delta t$ when the grain is located
at $X$ can be expressed as
\begin{equation}
  \mathcal{W}_\mathrm{L,R}^{(\pm)}(n,X,\Delta t)=\Delta t 
\frac{1}{R_\mathrm{L,R}(0)C}\Gamma_\mathrm{L,R}^{(\pm)}(n,X) \, ,
  \label{e.Wdef}
\end{equation}
where 
\begin{eqnarray}
  \Gamma_\mathrm{L}^{(\pm)}(n,X) &=& \left(\frac{R_\mathrm{L}(0)}
{R_\mathrm{L}(X)}\right)f(\pm\frac{{V}C}{2e}\mp n-\frac{1}{2}) \nonumber \\
   \Gamma_\mathrm{R}^{(\pm)}(n,X)&=& \left(\frac{R_\mathrm{R}(0)}
{R_\mathrm{R}(X)}\right)f(\pm\frac{{V}C}{2e}\pm n-\frac{1}{2}) \, .
 \label{e.Gammadef}
\end{eqnarray}
Here the function $f$ is defined as
\begin{equation}
 f(x)=\frac{x}{1-\exp({-\beta U_cx})} \, ,
 \label{e.fdef}
\end{equation}
where 
$R_\mathrm{L(R)}(X)$ is the tunneling resistance of 
the left (right) junction
which depends exponentially on the distance between the grain and the 
respective lead. In this paper 
$R_\mathrm{L}(X)=R_\mathrm{R}(-X)=R\exp(X/\lambda)$ where 
we refer to $\lambda$ as the {\em tunneling length}. Depending on the
material of the reservoirs and the insulating links $\lambda$ can be 
estimated to lie within $0.05{\rm{\AA}}-3{\rm{\AA}}$ for direct 
tunneling from the electrodes to the grain. Due to the strong exponential
dependence on the tunneling resistances the variations 
in capacitance with position are relatively small and 
are therefore neglected. 
The equations (\ref{e.EOM})-(\ref{e.fdef}) hence define our model system.
\section{Limit of Weak Electromechanical Coupling}
In this Section we describe how the model system introduced above can be
 solved analytically for the case of weak electromechanical
coupling. In later Sections this analytical 
solution will be compared to `exact' Monte-Carlo results and found to be
very useful in the further analysis of the model system.
%
%
The key is to average over the fast grain oscillations so that we are left
with a set of equations describing only the slow variations in 
the amplitude. 
We start by considering the typical scales of our parameters.
It is natural to use 
$\lambda$ as a characteristic length scale. Furthermore the condition for 
observing
Coulomb blockade effects requires that we operate with voltages ${V}$ 
of the order
of the Coulomb blockade offset voltage ${V}_0=U_c/e=e/C$. 
Introducing the 
dimensionless variables $x\equiv X/\lambda$ and 
${v}\equiv{V}/{V}_0$
the equation of motion for the grain (\ref{e.EOM}) turns into
\begin{equation}
 \ddot{x}+\omega^2x=\frac{\Omega^2}{2}{v}n(t)-\gamma\dot{x} \, ,
 \label{e.2}
\end{equation}
where $\omega=\sqrt{k/M}$ is the elastic oscillation frequency 
and $\gamma$ and 
$\Omega^2$ are defined through $\gamma\equiv\gamma_{d}/M$ and
\begin{displaymath}
\Omega^2\equiv\frac{e^2}{MLC\lambda}=\frac{\lambda}{L}
\frac{{U}_{c}}{E_\mathrm{\lambda}}\omega^2 \, .
\end{displaymath}
Here $E_\mathrm{\lambda}=k\lambda^2/2$ is the energy of harmonic mechanical 
vibrations with amplitude $\lambda$. For a nanoscale grain and typical organic 
junctions $\Omega^2/\omega^2=\epsilon\sim 10^{-2}$. Taking 
$\gamma/\omega$ to be of the same order\footnote{
Although the analytical model presented in this paper 
is dependent on treating the electromechanical coupling 
and the damping as small perturbations to a simple 
harmonic oscillator, computer simulations reveal that the 
qualitative behavior of the system
persists even when these terms are large.}
we can separate out slow variations
in oscillation amplitude by averaging over the fast oscillations.
This is conveniently done by
looking for a solution to (\ref{e.2}) of the form
\begin{equation}
x(t)=\tilde{x}(t)\sin(\omega t) \, ,
\label{e.3}
\end{equation}
where $\tilde{x}(t)$ is a slowly varying function such that 
$\dot{\tilde{x}}(t)\sim\epsilon\omega$.
Substituting (\ref{e.3}) into (\ref{e.2}), multiplying by
$\cos(\omega t)$ and averaging over a time interval 
$T$ yields
\begin{equation}
  \frac{\mathrm{d}\tilde{x}}{\mathrm{d}t}=\frac{\Omega^2}
{2\omega}{v}\left\langle n(t)\cos(\omega t)\right\rangle_{\tilde{x}(t)}
-\frac{1}{2}\gamma\tilde{x}(t) \, .
  \label{e.xtilde}
\end{equation}
The time average is indicated by brackets and defined as
\begin{displaymath}
  \left\langle g(t)\right\rangle_{\tilde{x}}=\frac{1}{T}
\int_{t-T/2}^{t+T/2}g(\tau)\mathrm{d}\tau \, ,
\end{displaymath}
where the index $\tilde{x}$ indicates that 
the averaging process is performed for
harmonic 
oscillations with {\em constant amplitude} $\tilde{x}$.  
The value of $T$ has to be chosen to obey the double inequality 
\begin{displaymath}
 \omega^{-1},\omega_\mathrm{R}^{-1}=RC \ll T \ll t_\mathrm{S}=\omega^{-1}
\epsilon^{-1}  \, ,
 \label{e.ineq}
\end{displaymath}
where $t_\mathrm{S}\sim 10^{-9}-10^{-8}\mathrm{ s}$ is the characteristic time 
for changes in the vibration amplitude $\tilde{x}(t)$ and 
$\omega_\mathrm{R}^{-1}\sim 10^{-10}\mathrm{ s}$ 
is the typical
time for charge redistribution between the leads and the grain. 
With this choice of $T$
the grain will 
perform many oscillations which differ very little in amplitude during the 
averaging, i.e.
\begin{displaymath}
\frac{\tilde{x}(t+T)-\tilde{x}(t)}{\tilde{x}(t)}\ll 1 \, .
\end{displaymath}
%

Introducing the dimensionless mechanical vibration energy 
\begin{displaymath}
 E(t)=\frac{M\omega^2(\lambda\tilde{x}(t))^2}{2E_\mathrm{\lambda}}
\end{displaymath}
and making a change of variables in (\ref{e.xtilde}) we get
\begin{equation}
\frac{\mathrm{d}E}{\mathrm{d}t}={v}\frac{\Omega^2}{\omega}W(E)-\gamma E
\label{e.dEdt}
\end{equation}
\begin{equation}
W(E)=\sqrt{E}\left\langle n(t)\cos(\omega t)\right\rangle_{E} \, .
 \label{e.WEdef}
\end{equation}
The physical meaning of the term containing $W(E)$ is that energy is 
pumped into the system
when the grain is oscillating. Due to the correlation between the charge 
fluctuations
$n(t)$ and the position of the grain, $x(t)=\sqrt{E(t)}\sin(\omega t)$, $W(E)$ 
will be nonzero. In order to describe this correlation we introduce the 
correlation function
\begin{equation}
 P_{n}(\varphi,E)=\pi\left\langle\delta_{n,n(t)}\delta
\left(\sin\half(\omega t-\varphi)\right)\right\rangle_{E} \, ,
 \label{e.Pndef}
\end{equation}
where the average is taken over a time $T$ that allows the grain to 
perform $N=T\omega/2\pi$ complete harmonic oscillations with constant energy
$E$. The grain will now pass through the particular point 
\begin{displaymath}
(E,\varphi)=(x,\dot{x})=(x_{E}(\varphi),\dot{x}_E(\varphi))=\left(\sqrt{E}\sin(\varphi),
\omega\sqrt{E}\cos(\varphi)\right)
\end{displaymath}
in a two-dimensional `phase space'
$N$ times and $P_n(\varphi,E)$ 
will thus be the relative number
of times the grain passes this point with charge $Q=en$. 
Using the definition (\ref{e.Pndef}) of $P_n(\varphi,E)$
we rewrite (\ref{e.WEdef}) as
\begin{equation}
 W(E)=\frac{\sqrt{E}}{2\pi}\int\mathrm{d}\varphi\cos(\varphi)q(\varphi,E) 
\label{e.WE}
\end{equation}
\begin{equation}
q(\varphi,E)\equiv\sum_n nP_n(\varphi,E) \, .
\label{e.qeq}
\end{equation}
In Appendix~\ref{A.a} it is shown how one can obtain a 
differential equation for 
$P_n(\varphi,E)$. The equations (\ref{e.dEdt}), 
(\ref{e.WE}) and (\ref{e.PnODE}) then completely describe
the behavior of the model and are stated below in their final form,
\begin{eqnarray}
\frac{\mathrm{d}E}{\mathrm{d}t}&=&{v}\frac{\Omega^2}{\omega}W(E)-\gamma E 
\label{e.sh-eq1}\\
 W(E) &=& \frac{\sqrt{E}}{2\pi}\int\mathrm{d}\varphi\cos(\varphi)\sum_n 
nP_n(\varphi,E) \label{e.sh-eq2} \\
\frac{\d{\vec{P}}(\varphi,E)}{\d\varphi} &=&\nu_\mathrm{R}
\hat{\mathcal{G}}(\sqrt{E}\sin\varphi)\vec{P}(\varphi,E)  \, .
 \label{e.sh-eq3}
\end{eqnarray}
Here $\vec{P}(\varphi,E)$ is a vector containing $P_n(\varphi,E)$ and
the components of the matrix 
$\hat\mathcal{G}$ appearing in (\ref{e.sh-eq3}) are
\begin{displaymath}
  \hat\mathcal{G}_{n,m}(x_E(\varphi))
=-\delta_{n,m}[\Gamma^+_E(n,\varphi)+\Gamma^-_E(n,\varphi)]
+\delta_{n,m\pm1}\Gamma^\mp_E(n,\varphi)
\end{displaymath}
\begin{displaymath}
\Gamma_\mathrm{E}^\pm(n,\varphi)=\Gamma_\mathrm{L}^\pm(n,x_E(\varphi))+
\Gamma_\mathrm{R}^\mp(n,x_E(\varphi)) 
\end{displaymath}
and we have defined a dimensionless charge relaxation frequency,
\begin{displaymath}
\nu_\mathrm{R}=\frac{\omega_\mathrm{R}}{\omega} \, .
\end{displaymath} 

If we formally solve equations (\ref{e.sh-eq1}-\ref{e.sh-eq3})
any observable characterizing the system can be evaluated. 
One such observable is
the average current through the left and rights leads respectively.
In Appendix \ref{A.a} this current is shown to be
\begin{equation}
\bar{I}_\mathrm{L,R}=\frac{e\omega}{2\pi}
\int_0^{2\pi}\mathrm{d}\varphi\sum_n
j_n^\mathrm{L,R}
(\sqrt{E}\sin\varphi)P_n(\varphi,E) \, ,
\label{e.Ibareq}
\end{equation}
where
\begin{eqnarray}
j_{n}^\mathrm{L,R}(\sqrt{E}\sin\varphi)=\nu_\mathrm{R}\mathrm{e}^{\mp\sqrt{E}
\sin\varphi}\left[f({v}/2\mp n-1/2)-f(\pm n-{v}/2-1/2)\right] . \nonumber \\
\label{e.jndef}
\end{eqnarray}
By using the definition (\ref{e.jndef}) of the partial currents and
Eq. (\ref{e.qeq}) together with
(\ref{e.PnODE}) one comes to the expression 
\begin{equation}
 \frac{\d q(\varphi,E)}{\d\varphi}=\sum_n\left(j_n^\mathrm{L}
(\sqrt{E}\sin\varphi)-j_n^\mathrm{R}(\sqrt{E}\sin\varphi)\right)P_n(\varphi,E)
\,.
 \label{e.qODE}
\end{equation}
This form of our equations is especially useful 
when $v=(2n+1)$ at zero temperature, i.e. 
when the function $f(x)=x\theta(x)$ and when the voltage is chosen at the 
point where a new channel is about to switch on. The differential equation 
(\ref{e.qODE}) for $q$ then simplifies to
\begin{equation}
  \frac{\d q(\varphi,E)}{\d\varphi}=\nu_\mathrm{R}\left((1-{v})
\sinh(\sqrt{E}\sin\varphi)-2q(\varphi,E)\cosh(\sqrt{E}\sin\varphi)\right).
\label{e.qsimple}
\end{equation} 
Instead of $n$ coupled differential equations for $P_n$ there is now
only one for $q(\varphi,E)$.

We are now ready to apply our analytical solution of the model
Coulomb blockade system to an analysis of
the shuttle instability. This will be done in the next two Sections.

\section{Analytical and Numerical Analysis of the Shuttle Instability}

In this Section we will use both the approximate analytical approach of the
previous Section and an `exact' numerical scheme to analyze the shuttle
instability in our model Coulomb blockade system. Using first the analytical
approach we conclude from Eq.~(\ref{e.sh-eq1})
that the stationary regimes of our system are defined by
\begin{displaymath}
v\frac{\Omega^2}{\omega}W(E)=\gamma E \, .
\end{displaymath}
This equation always has one trivial solution $E=0$. When this solution is 
stable it corresponds to the system being close to the point of mechanical 
equilibrium
only subject to small deviations due to charge fluctuations. We will refer 
to this
regime as the {\em static regime}. For our symmetric system the current-voltage
characteristics in this
regime show no pronounced Coulomb blockade structure 
even though there are peaks in the 
differential conductance due to the switching on of new channels at 
voltages $v_n=2n+1$ \cite{one}.

\begin{figure}[htbp]
\begin{center}\leavevmode
\includegraphics[width=0.6\linewidth]{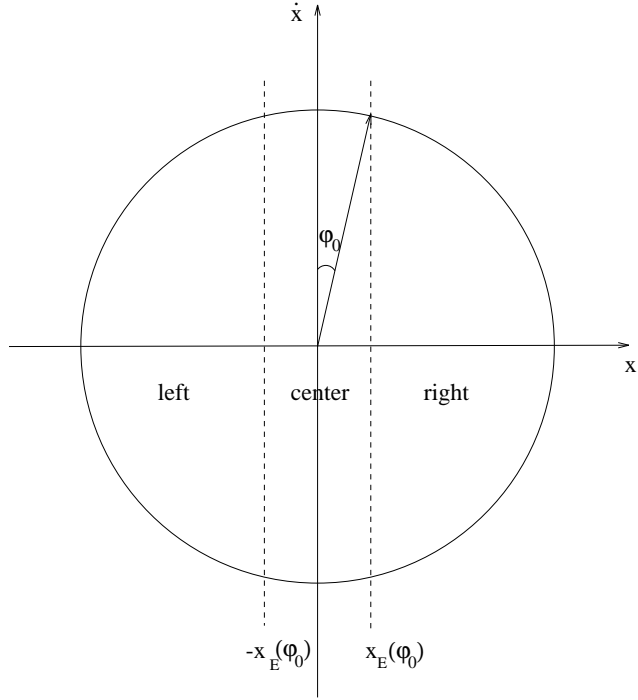}
\bigskip
\caption{Division of the phase space into three regions: 
 `left', `center' and  `right'. In the left and right regions the 
charge distribution on the grain is determined
by the exchange with the corresponding lead. In the center region
charge exchange with both leads contribute to the charge distribution. 
}\label{f.regions}\end{center}
\end{figure}

In Appendix~\ref{a.b} it is shown that for small $E$ we have
\begin{displaymath}
 W(E) = \alpha(v) {E}+\mathcal{O}(E^2),\mbox{ } \alpha(v)>0 \, .
\end{displaymath}
Hence, according to (\ref{e.sh-eq1}) the point $(x=0,\dot{x}=0)$ 
will become unstable if 
\begin{displaymath}
\frac{\d E}{\d t}\Bigg|_{E=0}>0 \mbox{ }\Leftrightarrow\mbox{ }
v\frac{\Omega^2}{\omega}\alpha(v)>\gamma \, ,
\end{displaymath}
i.e. when more energy is pumped into the system than can be dissipated.
We define the {\em critical voltage} $v_c$ as the voltage when
\begin{displaymath}
 v_c\frac{\Omega^2}{\omega}\alpha(v_c)=\gamma \, .
\end{displaymath} 
This equation cannot be solved for $v_c$ in the general case, but if
we know $\alpha(v)$ for some specific values $v_1$ and $v_2$ of the bias voltage
and if 
\begin{displaymath}
  v_1\frac{\Omega^2}{\omega}\alpha(v_1)<
\gamma<v_{2}\frac{\Omega^2}{\omega}\alpha(v_{2}) \, ,
\end{displaymath}
it follows that $v_c$ lies in the interval $[v_1,v_2]$
since $\alpha(v)$ is a continuous function of $v$.  
For the special case $v=v_n\equiv(2n+1)$ at zero temperature 
we only need to consider the single equation (\ref{e.qsimple}). By
performing successive partial integrations of (\ref{e.sh-eq2}) 
using (\ref{e.qsimple}) and approximating
 $\sinh{x}\approx x$ and $\cosh x\approx 1$
one finds that 
\begin{displaymath}
\alpha(v_n)=\frac{\nu_\mathrm{R}(v_n-1)}{2(1+4\nu_{R}^2)} \, .
\end{displaymath}
Hence the critical voltage can be established 
within one Coulomb blockade voltage,
\begin{equation}
    \left|v_c-2\left[\frac{3}{4}+\sqrt{\frac{\gamma\omega(1+4\nu_\mathrm{R}^2)}
{2\nu_\mathrm{R}\Omega^2}+\frac{1}{16}}\right] \right|<1 \label{e.Vcstep}
\end{equation}
If this critical voltage is exceeded the oscillation amplitude 
will increase as the voltage is raised
above $v_c$. The answer to the natural question whether this increase of  
amplitude will saturate at some definite value or not is determined by the 
behavior of $W(E)$
as $E$ tends to infinity.
To investigate the behavior of $W(E)$ for large values of 
$E$ it is convenient to 
consider the contributions 
to $W(E)$ from three different regions in $(x,\dot{x})$- space; a ``right'', 
a ``left'' and a ``central'' region, see Fig.~\ref{f.regions}. 
The ``right'' region
is defined as $x_E(\varphi)>x_E(\varphi_0)\equiv\ln E$, where the rate of the 
charge exchange between the grain and the right lead dominates over the  
exchange with the left lead. 
Similarly the ``left'' region is
defined as $x_E(\varphi)<-x_E(\varphi_0)$ where the converse is true. 
The third region is 
the central region where $|x_E(\varphi)|<x_E(\varphi_0)$. In a symmetric 
system the first 
two regions contribute equally to $W(E)$ and we need only consider the 
``left'' region.
For this case it is convenient to represent Eq.~(\ref{e.sh-eq3}) for
$\vec{P}(\varphi,E)$ in the form
\begin{equation}
\hat\mathcal{G}_\mathrm{L}\vec{P}=-\varepsilon(E)\tau_\mathrm{R}g(\varphi)
\frac{\d \vec{P}}{\d \varphi}
-\varepsilon^2(E)g^2(\varphi)\hat\mathcal{G}_\mathrm{R}\vec{P} \, ,
\label{e.Esatpert}
\end{equation}
where 
\begin{displaymath}
\hat\mathcal{G}_\mathrm{L,R}=e^{\pm x}\hat\mathcal{G}(x)|_{x\gg1},\mbox{ }
\varepsilon(E)=\exp(-x_E(\varphi_0))=E^{-1}
\end{displaymath}
\begin{displaymath}
g(\varphi)=\exp(\sqrt{E}(\sin\varphi_0+\sin\varphi))<1 \, ,
\end{displaymath}
i.e. $\hat\mathcal{G}_\mathrm{L,R}$ describes the charge exchange with 
the corresponding lead when the grain is disconnected from the
other lead. 
For large $E$ when $\varepsilon\ll1$ one can develop a perturbation procedure
to solve (\ref{e.Esatpert}). For the `left' region one finds the expansion
\begin{displaymath}
\vec{P}(\varphi)=\vec{P}^\mathrm{L}-
\varepsilon^2(E)g^2\hat\mathcal{G}_\mathrm{L}^{-1}\hat\mathcal{G}_\mathrm{R}
\vec{P}^\mathrm{L}+... \, ,
\end{displaymath}
where $\vec{P}^\mathrm{L}$, which satisfies the equation $\hat\mathcal{G}^
\mathrm{L}\vec{P}^\mathrm{L}=0$,
describes the charge distribution for a grain in thermal 
equilibrium
with the left lead. Therefore the charge on the grain will saturate at the 
value 
$q^\mathrm{L}=\sum n P_n^\mathrm{L}$ and as a consequence the leading 
term does not
depend on the position of the grain. This is why the contribution from the
left (and right) region to $W(E)$ will decrase with increasing amplitude. The
contribution from the central region is restricted by the upper limit
$4x_E(\varphi_0)\max\{q(\varphi)\}$. All the above considerations yield
the inequality 
\begin{displaymath}
W(E)<A\ln E+B \quad \mbox{when} \quad E\gg1 \, .
\end{displaymath} 
This inequality implies that at large amplitude the rate of dissipation, 
$\gamma E$, always exceeds the rate at which energy is pumped 
into the system leading
to a final stationary state with the grain oscillating with a finite 
amplitude $E=E_\mathrm{sh}$ if $v>v_c$. This oscillating
regime we refer to as the {\em shuttle regime}.
A more sophisticated treatment reveals that
\begin{displaymath}
W(E)=2q^\mathrm{L}\ln E+\mathcal{O}(1)\mbox{  when  }E\gg1 \, .
\end{displaymath}
This leads to the following estimation for $E_\mathrm{sh}$ 
as a function of the bias voltage at $v=v_c$ at zero 
temperature when 2$q^\mathrm{L}\sim v$
\begin{displaymath}
E_\mathrm{sh}=\frac{\Omega^2v^2}{\omega\gamma}\ln\left(\frac{\Omega^2v^2} 
{\omega\gamma}\right) \, .
\end{displaymath}

\begin{figure}[htbp]
\begin{center}\leavevmode
\includegraphics[width=0.8\linewidth]{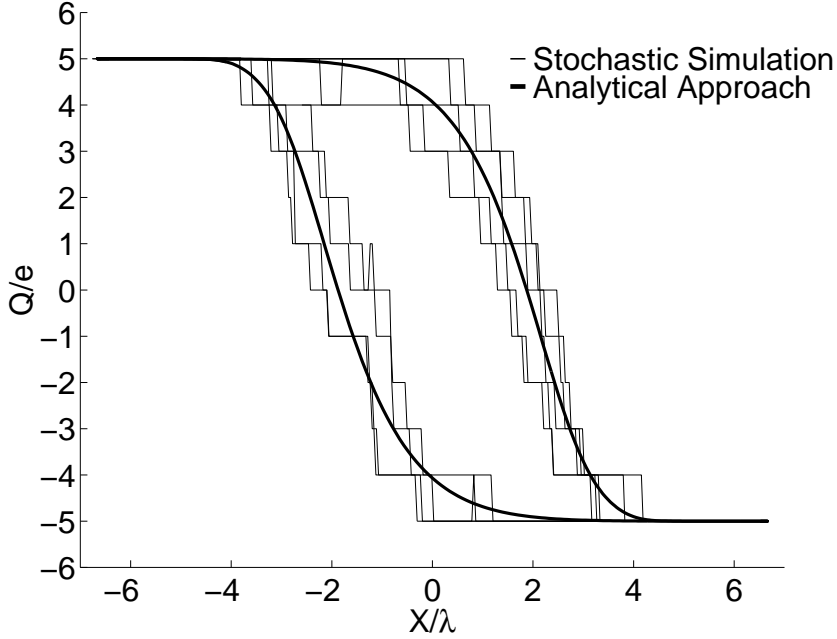}
\bigskip
\caption{Comparison between `exact' Monte Carlo simulations and the
analytical approach. The graph shows the grain charge
as a function of grain position in the
{\em shuttle regime} at bias voltage $V=10 V_0$ ($V_0$ is the theshold
voltage for lifting the Coulomb blockade). The thin
lines are from stochastic simulations and the thick 
smooth curve is the result of our analytical
approach to the system. From this graph we see that
the charging-decharging process takes place in
the center of the system while the the charge on
the grain saturates at it's maximum (minimum) value
outside this region. 
}\label{f.analstoch}\end{center}\end{figure}

\begin{figure}[htbp]
\begin{center}\leavevmode
\includegraphics[width=0.55\linewidth]{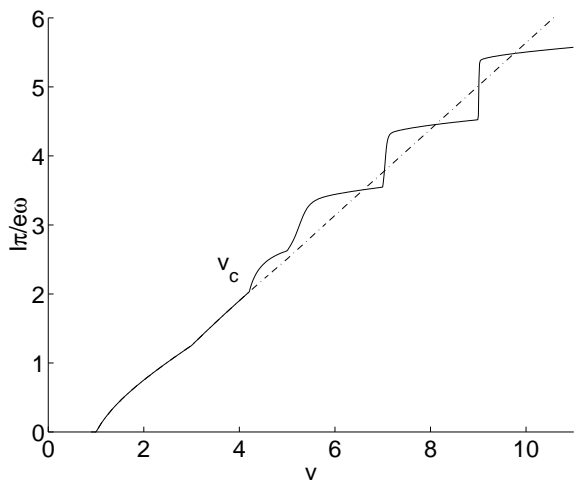}
\bigskip
\caption{Current through the Coulomb blockade system in the static regime
($V<V_c$) and in the shuttle regime ($V>V_c$). As the system enters the shuttle
regime the current (solid line) deviates from
the current for a static double junction (dashed line).
After the transition to the shuttle regime
distinct steps as predicted by
Eq.~(\ref{e.Ibarsh}) can be seen in the current even though we are modelling
a symmetric system. The current in this figure and subsequent ones is 
normalized to the frequency of harmonic oscillations to demonstrate 
that the step height in the shuttle regime is proportional to $\omega$.
In order to make a comparison with the current in the static regime this
current has been scaled by the same factor.   
}\label{f.IV}\end{center}\end{figure}

To support the preceeding discussion we have performed numerical
simulations of the system.  Figure~\ref{f.analstoch} shows the charge
on the grain as a function of its position in the 
shuttle regime.  
The thin lines are the result of a Monte-Carlo simulation of
the system when $v=10$ and the thick lines show the results
of our analytical approach. The graph reveals the
true stochastic nature of the charge on the grain
as well as the validity of our analytical solution.
Moreover one sees that the charging and decharging
of the system takes place in a limited 
region around the center of the system while the charge 
saturates as the grain approaches the respective lead.
To understand this behavior one can  
integrate Eq.~(\ref{e.qODE}) to 
get the following relationship for the charge on the grain
\begin{displaymath}
q(0)-q(\pi)=\int_{\pi}^{2\pi}\sum_nj_n^\mathrm{L}P_n(\varphi)\d\varphi
-\int_{\pi}^{2\pi}\sum_nj_n^\mathrm{R}P_n(\varphi)\d\varphi
\end{displaymath}
(Recall that $q(2\pi)=q(0)$ and that $q(0)$ [$q(\pi)$] is the charge on 
the grain as it passes the center position moving right [left]).
Substituting this expression
into (\ref{e.Ibareq}) one can get the current in the form
\begin{eqnarray}
\bar{I}=\frac{e\omega}{2\pi}[q(0)-q(\pi)]+
\frac{e\omega}{2\pi}\int_0^\pi\d\varphi\sum_nj_n^\mathrm{L}P_n(\varphi)+
\frac{e\omega}{2\pi}\int_\pi^{2\pi}\d\varphi\sum_nj_n^\mathrm{R}P_n(\varphi)\,.
\nonumber \\\label{e.Ibar}
\end{eqnarray} 
The last two terms determine the current between the grain
and the most distant lead. We can think of this contribution
as a {\em tunnel current} through
the central cross section of the system. The first term corresponds to
the current through this cross section when the grain passes the point
$x=0$. This current exists only because of the oscillatory motion 
of the grain. We refer to this mechanically mediated current as the 
{\em shuttle current}. At large amplitudes we can estimate the contribution
from the last two terms in (\ref{e.Ibar}) to be of the order $1/\sqrt{E}$.
At the same time it follows from a perturbative treatment analogous to the
one carried out above to determine the behavior of $W(E)$ for large $E$ that 
\begin{displaymath}
-q(\pi)=+q(0)=|q^\mathrm{L}|-\mathcal{O}\left(\frac{1}{\sqrt{E}}\right) \, .
\end{displaymath}  
At zero temperature we have also 
\begin{displaymath}
|q^\mathrm{L,R}|=\left[\frac{v+1}{2}\right] 
\end{displaymath}
where the square brackets $[..]$ denotes the integer part of the argument. Thus we find for this case that
\begin{equation}
\bar{I}=\frac{e\omega}{\pi}\left[\frac{v+1}{2}\right]+e\omega_\mathrm{R} 
\mathcal{O}\left(\frac{1}{\sqrt{E}}\right) \, .
\label{e.Ibarsh}
\end{equation}
From (\ref{e.Ibarsh}) one finds that for large amplitude
oscillations a pronounced steplike behavior in the current
appears in the symmetric case in contrast to the 
ordinary static Coulomb blockade case. This
behavior is also seen in numerical simulations of the system.
In Fig.~\ref{f.IV} the current-voltage characteristics
has been calculated. As the voltage is raised above $v_c$
the current departs from the current obtained in a static double
junction and distinct steps appear.

\section{Soft and Hard Excitation of Self-Oscillations}
We have in the previous Sections shown
that the stationary point $(x=0, \dot{x}=0)$ will become
unstable for certain voltages $v>v_c$ and that
the system will reach a self-oscillating 
regime with well defined amplitude. In this Section we now show 
that the transition from the {\em static regime} to the 
{\em shuttle regime} can be associated with either {\em soft} 
or {\em hard} excitation of self-oscillations. 
The terminology is adopted from the
theory of oscillatons \cite{Andronov}); in the case
of soft excitation of self-oscillations the amplitude increases smoothly
from zero at the transition point, while the oscillation amplitude jumps
to a finite value in the case of hard excitation of self-oscillations.
The former case occurs when $W^{\prime\prime}(0)<0$ and the latter
when $W^{\prime\prime}(0)>0$.
\begin{figure}[htbp]
\begin{center}\leavevmode
\includegraphics[width=0.55\linewidth]{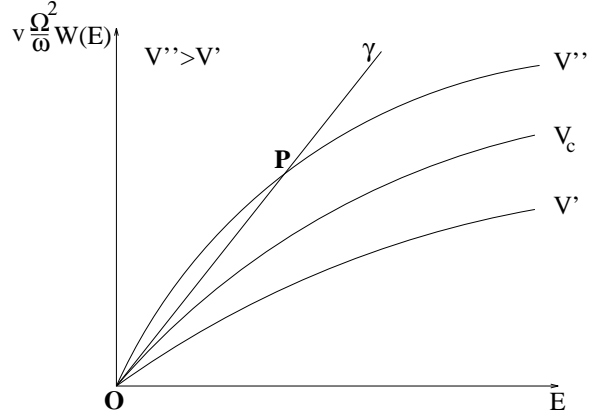}
\bigskip
\caption{Schematic energy diagram for the case $W^{\prime\prime}(0)<0$. 
The graph shows $W(E)$ for a fixed
set of parameters for three different voltages along with the line $\gamma E$.
When $v=v^\prime<v_c$ all the energy pumped into the system can be 
dissipated and the stationary point $\mathbf{O}$
is stable. When $v=v_c$ this point becomes unstable and at voltage $v=v'>v_c$ 
the system will reach a
limit cycle with oscillation amplitude $\propto\sqrt{E}$ determined by the 
intersection point $\mathbf{P}$.
}\label{f.mild}\end{center}\end{figure}

We start with the case when $W^{\prime\prime}(0)<0$
then for small $E$ and $v-v_c\ll v_c$ we have
\begin{displaymath}
W(E)=[\alpha(v_c)+\alpha^\prime(v_c)(v-v_c)]E-
\beta E^2 \mbox{ }\mbox{ } \alpha,\beta>0 \, .
\end{displaymath}
The energy at constant amplitude satisfies according to (\ref{e.sh-eq1})
\begin{displaymath}
 v\frac{\Omega^2}{\omega}[\alpha(v)-\beta E]=\gamma \, ,
\end{displaymath}
where $v_c\alpha(v_c){\Omega^2}/{\omega}=\gamma$.
Hence, in the vicinity of the transition the amplitude of oscillation will 
increase
smoothly as
\begin{equation}
 \tilde{x}=\sqrt{E}=\sqrt{\left(\frac{\alpha(v_c)+v_c\alpha^\prime(v_c)}{\beta}\right)
\frac{(v-v_c)}{v_c}} \, .
 \label{e.sqrt}
\end{equation}
The development of the instability can be understood from the
diagrams in Fig.~\ref{f.mild}. 
The graph shows $W(E)$ for a fixed
set of parameters for three different voltages along with the line $\gamma E$.
When $v<v_c$
the dissipation is larger than the pumping of energy into the system and the 
stationary point $\mathbf{O}$
is stable. When $v=v_c$ this point becomes unstable and at voltage $v>v_c$ 
the system will reach a
limit cycle with oscillation amplitude $\propto\sqrt{E}$ determined by the 
intersection point $\mathbf{P}$.
\begin{figure}[htbp]
\begin{center}\leavevmode
\includegraphics[width=0.55\linewidth]{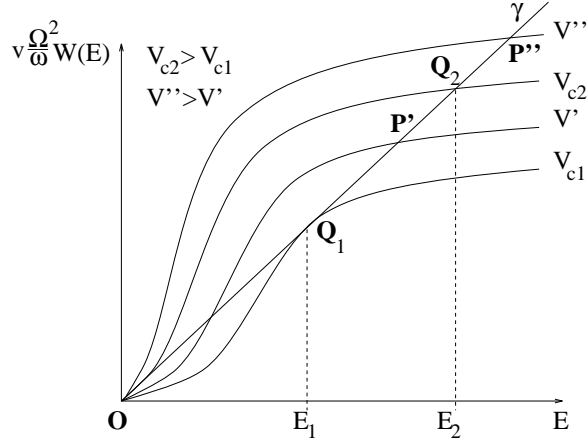}
\bigskip
\caption{Schematic energy diagram for the case $W^{\prime\prime}(0)>0$. 
The graph shows $W(E)$ for a fixed
set of parameters for four different voltages along with the line $\gamma E$. 
When $v<v_{c1}$
only $\mathbf{O}$ will be a stable stationary point. At $v=v_{c1}$ a 
second {\em unstable} stationary point $\mathbf{Q}_1$ appears. 
For $v_{c1}<v=v^\prime<v_{c2}$ we have two coexisting stable points 
$\mathbf{O}$ and $\mathbf{P^\prime}$
leading to the hysteretic behavior of the system discussed in the text.
At $v=v_{c2}$ $\mathbf{O}$ will become unstable and the system is determined 
to be in the limit
cycle with amplitude corresponding to energy $E_2$ at the intersection 
$\mathbf{Q}_2$.
As $v$ is increased above $v_{c2}$, the only stable stationary point left is
$\mathbf{P^{\prime\prime}}$ corresponding to a limit cycle with amplitude 
$\propto\sqrt{E}$.}
\label{f.hard}\end{center}\end{figure}

When $W^{\prime\prime}(0)>0$ the shuttle instability develops in a
completely different way. Consider the diagrams in Fig.~\ref{f.hard};
the graph shows $W(E)$ for a fixed
set of parameters for four different voltages along with the line $\gamma E$. 
Consider now the system being located in $\mathbf{O}$ at a voltage $v<v_{c1}$.
In this case the system is in the {\em static regime} and exhibits 
the same behavior as an ordinary double junction. As the voltage is increased above
$v_{c1}$ a second stable stationary point $\mathbf{P^\prime}$ appears 
but the system cannot reach this
point since $\mathbf{O}$ is still stable. At $v=v_{c2}$, $\mathbf{O}$
becomes unstable and the system ``jumps'' from $\mathbf{O}$ to $\mathbf{Q}_2$.
This instability we refer to as {\em hard} since the amplitude changes
abruptly from $E=0$ to $E=E_2$ as the voltage is raised above $v_{c2}$.
Now consider the case when the system is originally in the stationary point
$\mathbf{P^{\prime\prime}}$ and the voltage is lowered. At $v<v_{c2}$, 
$\mathbf{O}$ becomes stable but cannot be reached by the system until
$v$ has dropped to $v=v_{c1}$. At $v_{c1}$ the point $\mathbf{Q}_1$
becomes unstable and the system will ``jump'' to $\mathbf{O}$. 
This transition is characterized by an abrupt drop in amplitude
from $E_1$ to $E=0$ at $v=v_{c1}$. Since $v_{c1}<v_{c2}$ the system 
will obviously exhibit a hysteretic behavior
in the transition region.   

By using the simplified equation (\ref{e.qsimple}) for the charge
valid at zero temperature and $v=2n+1$ one can develop a perturbation
expansion for $q(\varphi)$ for small $E$. Solving the
system to third order in $\sqrt{E}$ we find an expression
for $W^{\prime\prime}(0)$
\begin{equation}
\frac{\d^2 W(E)}{\d E^2}\Bigg|_{E=0}=\frac{\tau_\mathrm{R}(v-1)}
{16(\tau_\mathrm{R}^2+4)^2}(\tau_\mathrm{R}^2-12) \, .
\label{e.derivcond}
\end{equation}
From this result follows that the instability is associated with
soft excitation of self-oscillations if
$\tau_\mathrm{R}=\nu_\mathrm{R}^{-1}<2\sqrt{3}$ and with 
hard excitation of self-oscillations if
$\tau_\mathrm{R}=\nu_\mathrm{R}^{-1}>2\sqrt{3}$.

\begin{figure}[htbp]
\begin{center}\leavevmode
(a)\includegraphics[width=0.4\linewidth]{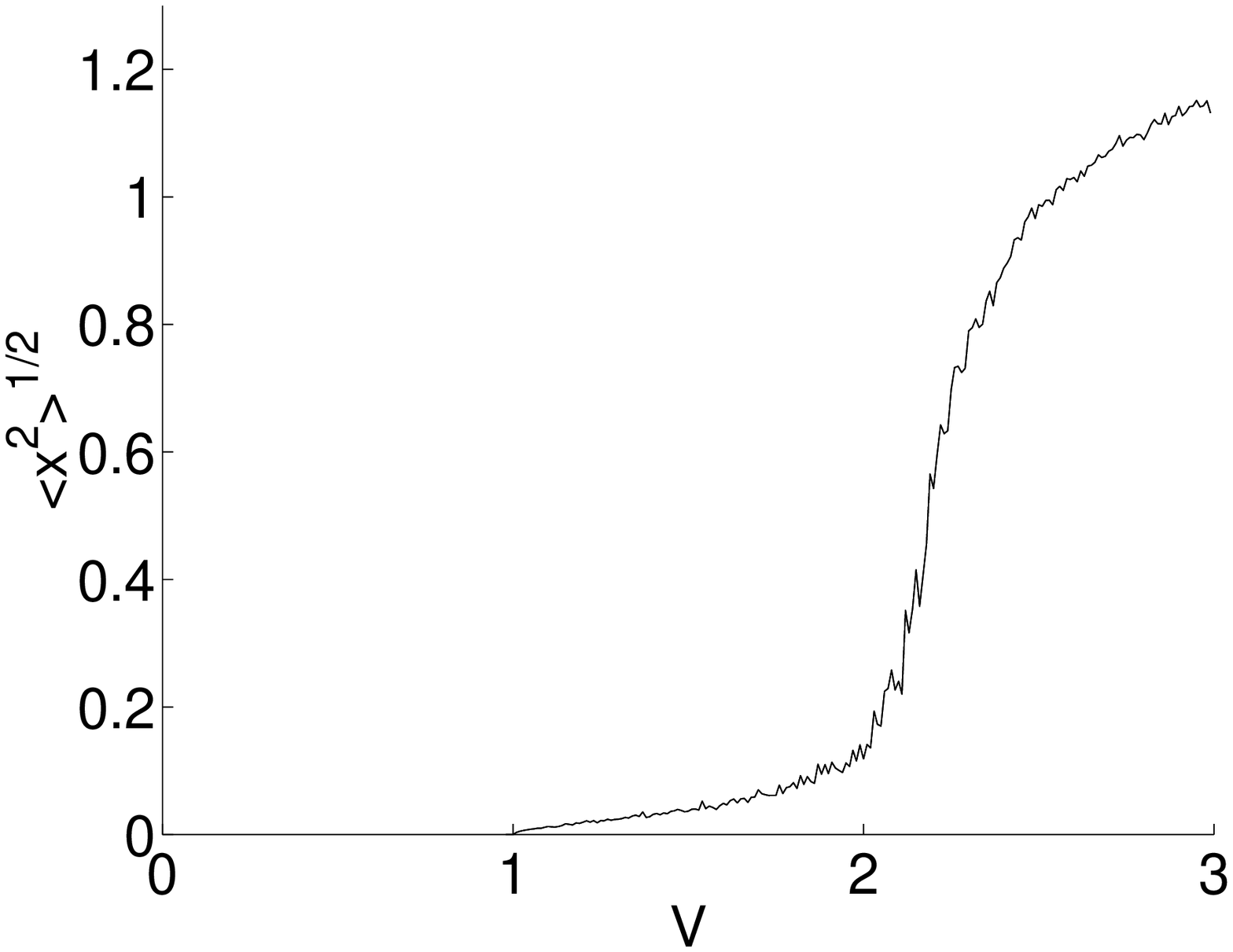}
(b)\includegraphics[width=0.4\linewidth]{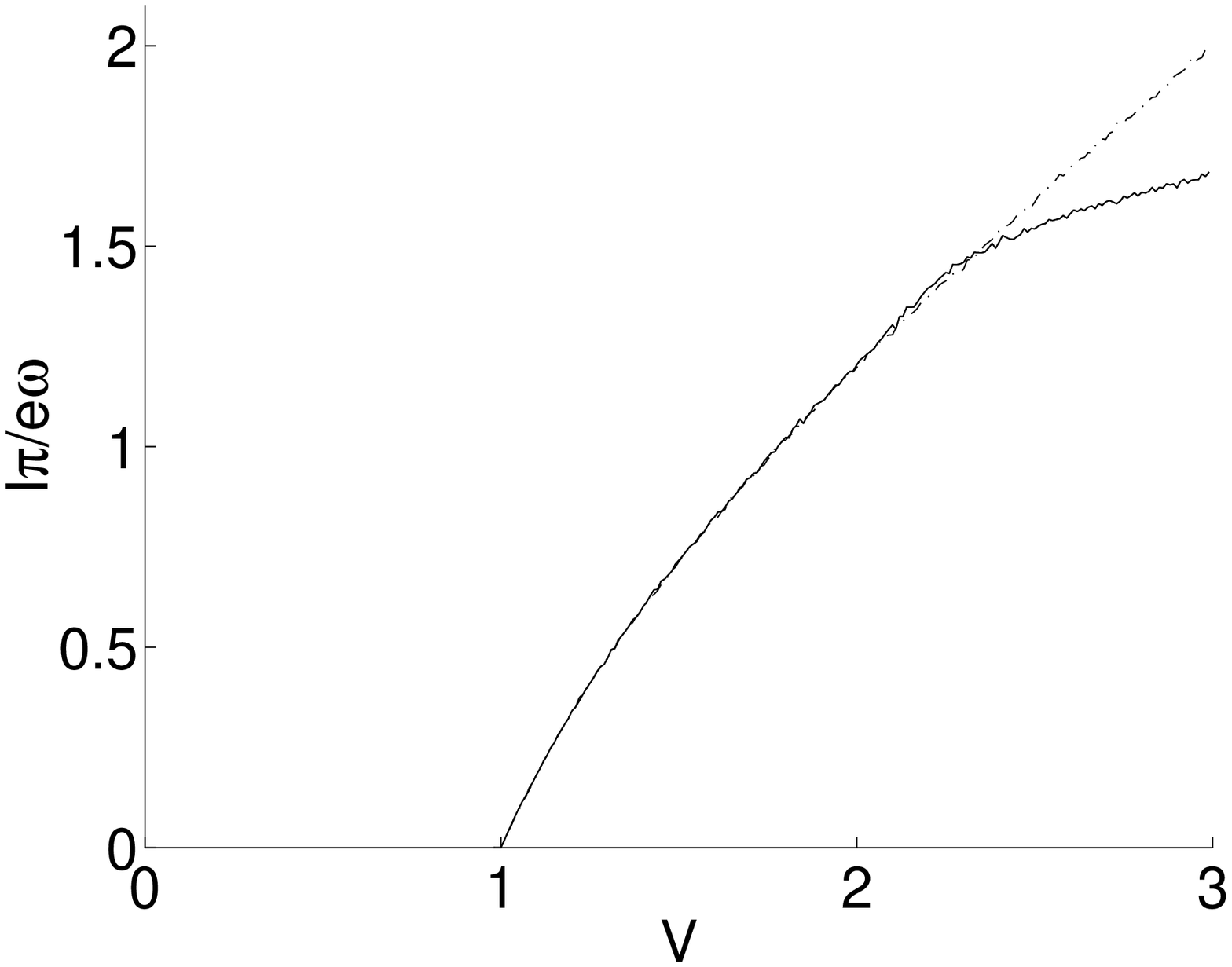}
\bigskip
\caption{Soft excitation of self-oscillations from Monte Carlo 
simulations. 
(a) shows the 
root mean square amplitude as the voltage is raised above $V_{c}$ when
$\tau_\mathrm{R}<2\sqrt{3}$. The predicted square-root increase of amplitude
(Cf. Eq.~(\ref{e.sqrt})) can clearly bee seen here. The slight increase in 
amplitude before the transition comes from fluctuations in the 
charge on the grain. (b) The loss of stability is reflected in the current
- voltage
characteristics. As the system enters the {\em shuttle regime} the current
deviates from the current seen in ordinary static double junctions 
(dashed line). 
}\label{f.MCmildtrans}\end{center}\end{figure}

Monte Carlo simulations of the system support the 
existence of two different types of instabilities. 
In Fig.~\ref{f.MCmildtrans} the result of a simulation for the case
when $\tau_\mathrm{R}<2\sqrt{3}$ is shown. 
As expected the amplitude grows as in (\ref{e.sqrt}). The transition
to the shuttle regime is visible in the current - voltage characteristics as
a lowering of the current compared to the current for a static
double junction. This decrease can be understood from the analytical
expression (\ref{e.Ibarsh}) for the current .

\begin{figure}[htbp]
\begin{center}\leavevmode
(a)\includegraphics[width=0.4\linewidth]{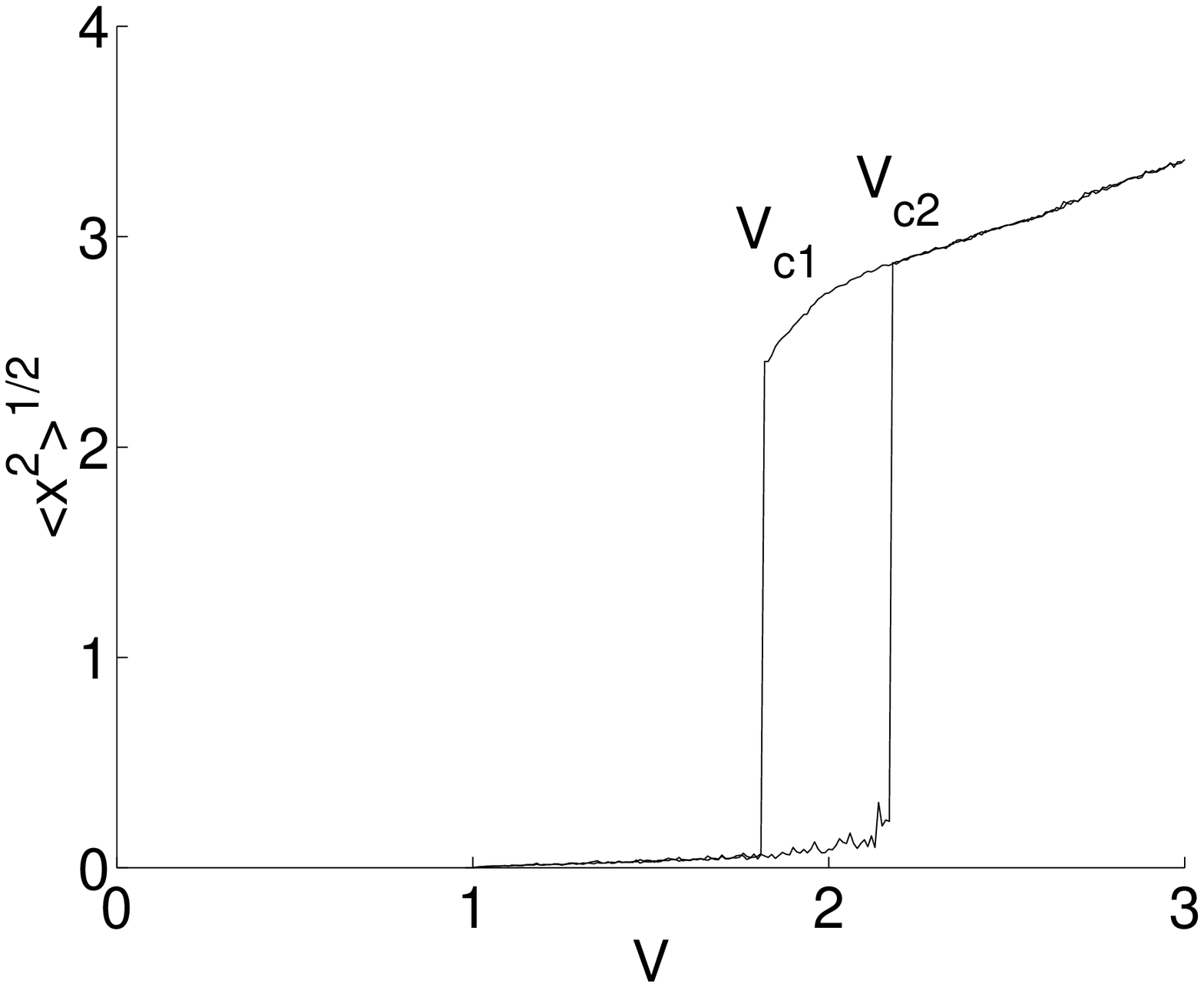}
(b)\includegraphics[width=0.4\linewidth]{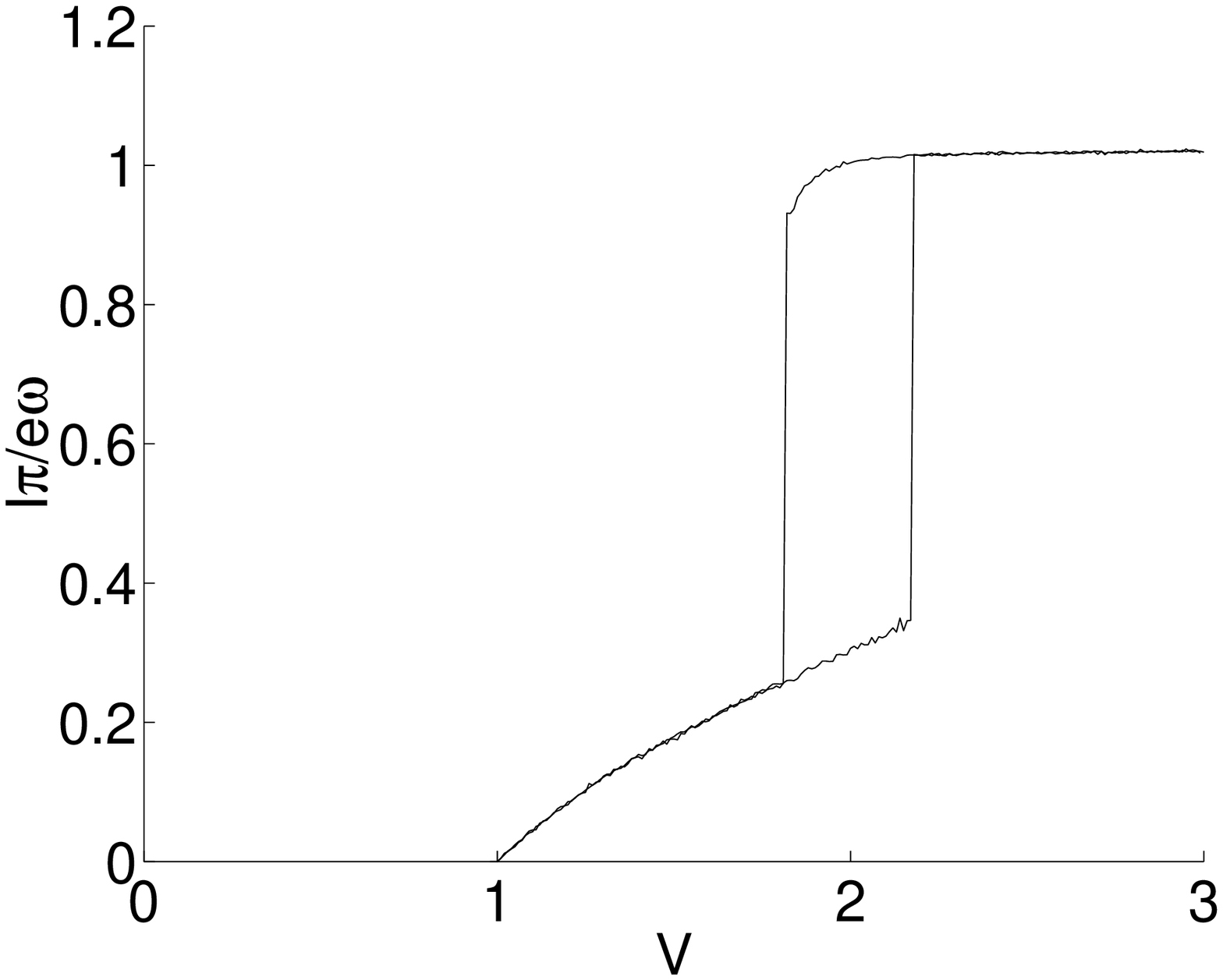}
\bigskip
\caption{Hard excitation of self-oscillations from Monte Carlo simulations. 
(a) shows
the root mean square amplitude at different voltages. As the voltage
is raised from $V=0$ there is a discrete jump in amplitude at $V=V_{c2}$.
Lowering the voltage again from $V=3V_0$ to zero 
reveals the hysteretic behavior 
as the amplitude drops to zero at $V=V_{c1}$. (b) The hysteretic behaviour
is clearly visible in the current. Since $\nu_\mathrm{R}\ll 1$ the current
in the {\em shuttle regime} lies very close to the value $I=e\omega/\pi$
as predicted in (\ref{e.Ibarsh}).    
}\label{f.MChardtrans}\end{center}\end{figure}

A similar simulation for a case when $\tau_\mathrm{R}>2\sqrt{3}$ is shown in
Fig.~\ref{f.MChardtrans}. 
The predicted hysteretic behaviour is clearly visible along with
the expected quantization of the current in the shuttle regime.

\section{Conclusions}

We have analyzed by both numerical and analytical methods charge transport
by a novel `shuttle mechanism' through the model shown in Fig.~1 of a 
self-assembled composite Coulomb blockade system \cite{prl}. A dynamical
instability was found to exist above a critical bias voltage $V_c$, which 
depends on the junction resistances in the system. In this `shuttle regime'
there is a limit cycle in the position-charge plane for the grain (shuttle)
motion as shown in Fig.~\ref{f.analstoch}. 
Above $V_c$ the current - voltage curve has a 
step-like  structure, a type of Coulomb staircase, as shown in 
Fig.~\ref{f.IV} even
though we are modelling a symmetric double-junction system.

The transition from the static regime, where the grain does not move, to 
the shuttle regime can either be associated with soft excitation of
self-oscillations, i.e.
 a continuous square-root increase in oscillation amplitude above $V_c$
as shown in Fig.~\ref{f.MCmildtrans} 
or by hard excitation of self-oscillations (the terminology is from the
theory of oscillations \cite{Andronov})
implying a sudden jump in oscillation amplitude at an upper critical voltage
$V_{c2}$. In the latter case the system behavior is
hysteretic, as shown in Fig.~\ref{f.MChardtrans}; when the bias
voltage is lowered the static regime is re-entered at a lower critical voltage
$V_{c1}$. 

Hard excitation of self-oscillation appears in our model for large junction
resistances, $\tau_R\equiv RC > 2\sqrt{3}/\omega$. At the same time
the numerical value of $V_c \propto \sqrt{R}$ 
for large resistance junctions and
may therefore be quite a bit larger than the threshold voltage $V_0$
for lifting the Coulomb blockade. The current in the
high-voltage shuttle regime may furthermore be distinctly larger than in
the static low-voltage regime. This is because it is governed mainly by 
the elastic vibration frequency of the grain and not by the rate of charge 
tunneling from the grain at rest in the center of the system, equally 
far from both leads. Hence in an experiment one might find a sudden increase
of current at a quite large voltage
$V_c$, where the shuttle mechanism sets in,
rather than at the lower threshold voltage $V_0$, where the Coulomb blockade 
is lifted. This 
observation may be relevant in connection with the recent experiment by
Braun {\em et al.} \cite{Sivan} where hysteresis effects in the current -
voltage curves and an anomalously large threshold voltage 
(if interpreted within
conventional Coulomb blockade theory) were reported for a system showing very
large resistance at low bias voltages. The results of this work show that it
may be possible to explain this experiment within a Coulomb blockade theory
which allows for a shuttle instability of the type described here.
In order to establish whether this mechanism is truly responsible for
the behaviour of the system in \cite{Sivan} one has to study 
the experimental situation in great detail to properly
estimate the mechanical parameters (elastic vibration frequency
and damping) of the insulating material. 

\begin{ack}
This work has been supported by grants from the Swedish Research Council
for Engineering Sciences, the Royal Swedish Academy of Science, 
and the Swedish Natural Science Research Council.
\end{ack}

\begin{appendix}
\section{Appendix A}
\label{A.a}
In this Appendix we derive the differential equation (\ref{e.sh-eq3})
for the quantity $P_n(\varphi,E)$ defined in (\ref{e.Pndef}) and the
result (\ref{e.Ibareq}) for the average current 
$\bar{I}_\mathrm{L,R}$through the left and right leads.

Denote by $N_n(\varphi,E)=NP_n(\varphi,E)\label{e.NnPn}$ 
the number of times the grain passes through the 
point $(E,\varphi)$
carrying charge $Q=en$ and let \\
$N_{mn}(\varphi,E,\Delta t)$ 
be the number of 
times
the grain has charge $Q=em$ as it passes through the point 
$(E,\varphi+\omega\Delta t)$ provided that
it passed through $(E,\varphi)$ at a time $\Delta t$ earlier with charge 
$Q=en$, i.e.
\begin{equation}
 N_{mn}(\varphi,E,\Delta t) = N\pi\left\langle 
\delta_{m,n(t+\Delta t)}\delta_{n,n(t)}\delta
\left(\sin\left(\half(\omega t-\varphi)\right)\right)\right\rangle_E
\end{equation}
It's now easy to see that 
\begin{eqnarray}
 N_n(\varphi+\omega\Delta t,E) &=& \sum_mN_{nm}(\varphi,E,\Delta t)=
\sum_{m\neq n}N_{nm}(\varphi,E,\Delta t)\nonumber \\
&+&\left(N_n(\varphi,E)-\sum_{m\neq n}N_{mn}(\varphi,E,\Delta t)\right)
\label{e.Neq}
\end{eqnarray}
Using the definition of $P_n(\varphi,E)$ in (\ref{e.Neq}) yields
\begin{eqnarray}
 P_n(\varphi+\omega\Delta t,E) &=& \sum_{m \neq n}\frac{N_{nm}
(\varphi,E,\Delta t)}
{N_m(\varphi,E)}P_{m}(\varphi,E) \nonumber \\
&+&\left(P_n(\varphi,E)-\sum_{m\neq n}\frac{N_{mn}(\varphi,E,\Delta t)}
{N_n(\varphi,E)}P_n(\varphi,E)\right)
\end{eqnarray}
For large $N$ the ratio $N_{nm}(\varphi,E,\Delta t)/N_m(\varphi,E)$ is the 
probability for the transition
$(n,Q_\mathrm{L,R})\rightarrow(m,Q_\mathrm{L,R}\mp (m-n))$ during a small 
time interval $\Delta t$. 
But the probability for this kind of event was given in (\ref{e.Wdef}) 
\begin{eqnarray}
P_n(\varphi+\Delta\varphi,E)&=&\left(1-\sum_{\sigma=\pm}
\sum_{\mathrm{S=L,R}}\mathcal{W}_\mathrm{S}^\sigma
\left(n,x_E(\varphi),\frac{\Delta\varphi}{\omega}\right)\right)
P_n(\varphi,E)\nonumber \\
&+&\sum_{\sigma=\pm}\sum_{\mathrm{S=L,R}}\mathcal{W}_\mathrm{S}^\sigma
\left(n-\sigma 1,x_E(\varphi),\frac{\Delta\varphi}{\omega}\right)
P_{n-\sigma 1}(\varphi,E)
\end{eqnarray} 
We can write this as a first order differential equation if we introduce 
\begin{eqnarray}
 \Gamma_\mathrm{E}^\pm(n,\varphi)&=&\Gamma_\mathrm{L}^\pm(n,x_E(\varphi))+
\Gamma_\mathrm{R}^\mp(n,x_E(\varphi)) \nonumber \\
&=&\mathrm{e}^{-\sqrt{E}\sin\varphi}f(\pm \frac{v}{2}\mp n-\frac{1}{2})+
\mathrm{e}^{\sqrt{E}\sin\varphi}f(\mp \frac{v}{2}\mp n-\frac{1}{2})
\end{eqnarray}
and use the relations (\ref{e.Wdef}) and (\ref{e.Gammadef})
\begin{eqnarray}
\frac{\mathrm{d}P_n}{\mathrm{d}\varphi}&=&-\frac{\omega_\mathrm{R}}{\omega}
\left(\Gamma_\mathrm{E}^+(n,\varphi)+\Gamma_\mathrm{E}^-(n,\varphi)\right)
P_n(\varphi,E)\nonumber \\
&+&\frac{\omega_\mathrm{R}}{\omega}
\left(\Gamma_\mathrm{E}^-(n+1,\varphi)P_{n+1}(\varphi,E)+\Gamma_\mathrm{E}^+
(n-1,\varphi)
P_{n-1}(\varphi,E)\right)
\label{e.PnODE}
\end{eqnarray}
The average current through the left (right) lead can formally
be written as
\begin{equation}
\bar{I}_\mathrm{L,R}=e\left\langle \frac{1}{\Delta t}\sum_{\sigma=\pm}
\sigma\mathcal{W}_\mathrm{L,R}^\sigma
\left(n(t),X(t),\Delta t\right)\right\rangle 
\end{equation}
Within our approximation we now consider this average when the grain is 
oscillating with a
fixed amplitude $\lambda\sqrt{E}$ and arrive at the equation
\begin{eqnarray}
\bar{I}_\mathrm{L,R}&=&\frac{e}{\Delta t}\int\mathrm{d}\varphi\sum_n
\sum_{\sigma=\pm}\sigma\mathcal{W}_\mathrm{L,R}^\sigma
\left(n,\lambda\sqrt{E}\sin\varphi,\Delta t\right)
\half\left\langle\delta_{n,n(t)}\delta(\sin\half(\omega t-\varphi))
\right\rangle \nonumber \\
\end{eqnarray}
By using the definition of the partial currents (\ref{e.jndef}) one 
gets eq.~(\ref{e.Ibareq}). 

\section{Appendix B}
\label{a.b}
In this Appendix we prove that the function $W(E)$ is a linear function
of $E$ for small $E$ with a positive coefficient.

For small $E$ we can expand (\ref{e.sh-eq3}) around $\sqrt{E}=0$ to obtain
\begin{equation}
 \frac{\d{\vec{P}}(\varphi,E)}{\d\varphi} =\nu_\mathrm{R}\left(
\hat{\mathcal{G}}_0+\sqrt{E}\sin\varphi\hat{\mathcal{G}_1}+...\right)
\vec{P}(\varphi,E) 
\label{e.sh3exp}
\end{equation} 
where
\begin{equation}
\hat{\mathcal{G}}_0=\hat{\mathcal{G}}(0) \mbox{ and } \hat{\mathcal{G}}_1
\equiv\frac{\d\hat{\mathcal{G}}(x)}{\d x}\Bigg{|}_{x=0}
\end{equation} 
Expanding $\vec{P}(\varphi,E)$ in the same fashion
\begin{equation}
\vec{P}(\varphi,E)=\vec{P}^{(0)}+\sqrt{E}\vec{P}^{(1)}(\varphi)+...
\label{e.Pansatz}
\end{equation}
and inserting this Ansatz into (\ref{e.sh3exp}) we find that $\vec{P}^{(0)}$ 
is the solution to the homogenous equation
\begin{equation}
  \hat{\mathcal{G}}_0\vec{P}^{(0)}=0
\end{equation}
The correction to first order in $\sqrt{E}$, $\vec{P}^{(1)}(\varphi)$ is the 
solution to
the equation
\begin{equation}
\left(\tau_\mathrm{R}\frac{\d}{\d\varphi}-\hat{\mathcal{G}}_0\right)
\vec{P}^{(1)}(\varphi)=\hat{\mathcal{G}}_1\vec{P}^{(0)}\sin\varphi
\end{equation}
this is a standard differential equation which can be solved to yield
\begin{eqnarray}
\vec{P}^{(1)}(\varphi)=\frac{
\nu_\mathrm{R}}
{\hat{I}+\nu_\mathrm{R}^2\hat{\mathcal{G}}_0^2}\hat\mathcal{G}_1\vec{P}^{(0)}
\cos\varphi+
\hat{\mathcal{G}}_0\frac{\nu_\mathrm{R}^2}
{\hat{I}+\nu_\mathrm{R}^2\hat{\mathcal{G}}_0^2}\hat\mathcal{G}_1\vec{P}^{(0)}
\sin\varphi
\end{eqnarray}
where $\hat{I}$ denotes the identity matrix and the matrix divisions are 
symbolic for the multiplication whith the inverse operator.
Recalling the expression (\ref{e.sh-eq2}) for $W(E)$ and introducing the 
vector $\hat{n}$ having components $n$ we find when inserting 
our expansion that we have to first order in $E$
\begin{eqnarray}
 W(E) &=& \frac{\sqrt{E}}{2\pi}\int\mathrm{d}\varphi\cos(\varphi)\hat{n}
\cdot \vec{P}(\varphi,E)\nonumber \\
&=& \frac{E}{2}\left(\hat{n}\cdot\frac{\nu_\mathrm{R}}
{\hat{I}+\nu_\mathrm{R}^2\hat{\mathcal{G}}_0^2}\hat\mathcal{G}_1
\vec{P}^{(0)}\right)+\mathcal{O}(E^2)
=\alpha {E}+\mathcal{O}(E^2)
\label{e.Walpha}
\end{eqnarray}
However, to get instability we require that $\alpha$ defined as
\begin{equation}
\alpha=\frac{1}{2\pi}\int\mathrm{d}\varphi\cos(\varphi)\sum_n 
nP_n^{(1)}(\varphi)
\end{equation}
must obey $\alpha>0$. To show this
consider the auxiliary equation
\begin{equation}
\left(\tau_\mathrm{R}\frac{\d}{\d\tau}-\hat{\mathcal{G}}_0\right)
\vec{P}^{(1)}(\tau)=\hat{\mathcal{G}}_1\vec{P}^{(0)}g(\tau)
\end{equation}
where $g(\tau)$ can be any non singular real valued function. We now define 
\begin{equation}
q^{(1)}(\tau)=\sum_nnP^{(1)}(\tau)
\end{equation}
and the functional $A\{g(\tau)\}$ as
\begin{equation}
A\{g(\tau)\}=\frac{\int_{-\infty}^\infty g^{\prime}(\tau)q^{(1)}(\tau)\d\tau}
{\int_{-\infty}^\infty g^{2}(\tau)\d\tau}
\label{e.Afunctional}
\end{equation}
Note that we have as a special case
\begin{equation}
\alpha=\frac{1}{2}{A\{g(\tau)=\sin\tau\}}
\end{equation}
From the theory of linear response  
\begin{equation}
q^{(1)}(\tau)=\int_{-\infty}^{\infty}\chi(\tau-\tau^{\prime})g(\tau^{\prime})
\d \tau^{\prime}
\label{e.qresponse}
\end{equation}
where $\chi(\tau-\tau^{\prime})$ is the response function. 
Using (\ref{e.qresponse}) in conjunction with (\ref{e.Afunctional}) one gets
\begin{equation}
A\{g(\tau)\}=\frac{\int\d\omega\mathrm{Im}\{\omega\chi(\omega)\}|g(\omega)|^2}
{\int\d\omega|f(\omega)|^2}
\end{equation}
where the Fourier transform is defined here as
\begin{equation}
g(\omega)=\int\d \tau\mathrm{e}^{\mathrm{i}\omega \tau}g(\tau)
\end{equation}
Causality together with the requirement that $q^{(1)}(\tau)$ be real give 
two conditions
\begin{equation}
\mathrm{Im}\{\chi(\omega)\}\neq 0 \mbox{ for } \omega\neq 0
\mbox{ and } 
\mathrm{Im}\{\chi(-\omega)\}=\mathrm{Im}\{-\chi(\omega)\} \nonumber
\end{equation}
Since $\chi$ is odd and nonzero at $\omega=0$ we conclude that
\begin{equation}
\mathrm{sgn}\left(\mathrm{Im}\{\omega\chi(\omega)\}\right)=Const
\end{equation}
However, the response function $\chi(\omega)$ is independent of the particular
 form of $g(\tau)$ and hence we have that
\begin{equation}
\mathrm{sgn}(\alpha)=\mathrm{sgn}\left(A\{\sin\tau\}\right)=\mathrm{sgn}
\left(A\{g(\tau)\}\right)
\end{equation}
To see that $\alpha>0$ consider the special choice of $g(\tau)$
\begin{eqnarray}
   g_0(\tau)=\left\{
       \begin{array}{cc}
           {0} & {|\tau|>T}\\
           {-g_0\mathrm{sgn}(\tau)} & {|\tau|<T}
       \end{array}\right.
\end{eqnarray}
where $T\gg{\omega}/{\omega_\mathrm{R}}$. This choice of $g$ has a very 
simple physical meaning. 
At $\tau<-T$ the grain is immobile in the centre of the system carrying charge 
$q=0$. At $\tau=-T$ it 
instantanously jumps to a location close to the right lead and waits 
there long enough to 
acquire a negative charge $q=-q_0$. It then moves instantanously to the left 
lead and there gets the
charge $q=+q_0$ and then moves back to the middle. 
This process yields
\begin{equation}
 A\{g_0(\tau)\}=3q_0/2T>0 
\end{equation}
Hence we have shown that for small $E$
\begin{equation}
 W(E) = \alpha {E}+\mathcal{O}(E^2),\mbox{ }\alpha>0
\end{equation}
\end{appendix}


\begin{thebibliography}{9}
\bibitem{one}
I. O. Kulik and R. I. Shekhter, Sov. Phys. JETP ${\bf 41}$, 308 (1975).

\bibitem{L}
D. V. Averin and K. K. Likharev, in {\em Mesoscopic Phenomena in Solids},
edited by B. L. Altshuler, P. A. Lee, and R. A. Webb (Elsevier, 
Amsterdam, 1991),
p.173.

\bibitem{Sivan}
E. Braun, Y. Eichen, U. Sivan, and G. Ben-Yoseph,
Nature {\bf 391}, 775 (1998).

\bibitem{two}
R. P. Andres, T. Bein, M. Dorogi, S. Feng, J. I. Henderson, C. P. Kubiak,
W. Mahoney, R. G. Osifchin, and R. Reifenberger, 
Science ${\bf 272}$, 1323 (1996).

\bibitem{three}
D. L. Klein, J. E. B. Katari, R. Roth, A. P. Alivisatos, and P. L. McEuen
Appl. Phys. Lett. ${\bf 68}$, 2574 (1996).

\bibitem{four}
E. S. Soldatov, V. V. Khanin, A. S. Trifonov, S. P. Gubin, V. V. Kolesov,
D. E. Presnov, S. A. Iakovenko, and G. B. Khomutov,
JETP Lett. {\bf 64}, 556 (1996).

\bibitem{Mm}
O. Alvarez and R. Lattore Biophys.J. {\bf 21}, 1, (1978);
V. I. Pasechnic and T. Gyanik, Sov. Biophyzika {\bf 22}, 941 (1977).

\bibitem{prl}
L. Y. Gorelik, A. Isacsson, M. V. Voinova, B. Kasemo, R. I. Shekhter,
and M. Jonson,
Phys. Rev. Lett. {\bf 80}, 4526 (1998); Physica B (in press) (1998).

\bibitem{Andronov}
A. A. Andronov, A. A. Vitt, and S. E. Khaikin,
{\it Theory of Oscillators} (Pergamon, Oxford, 1966), Ch.~IX, \S9.


 


\end{thebibliography}
\end{document}